\documentclass[12pt, reqno]{amsart}
\usepackage{amsmath, amsthm, amscd, amsfonts, amssymb, graphicx, color}
\usepackage[bookmarksnumbered, colorlinks, plainpages]{hyperref}
\hypersetup{colorlinks=true,linkcolor=red, anchorcolor=green, citecolor=cyan, urlcolor=red, filecolor=magenta, pdftoolbar=true}
\usepackage{float}
\theoremstyle{definition}

\usepackage{multirow}
\usepackage{stackengine}

\numberwithin{equation}{section}
\usepackage{pdflscape}

\usepackage{natbib}
\usepackage{tikz}
\usepackage{algorithm}
\usepackage{algpseudocode}
\usetikzlibrary{shapes.geometric, arrows}

\tikzstyle{startstop} = [rectangle, rounded corners, minimum width=2cm, minimum height=1cm,text centered, draw=black, fill=red!30]
\tikzstyle{process} = [rectangle, minimum width=2cm, minimum height=1cm, text centered, draw=black, fill=blue!30]
\tikzstyle{decision} = [circle, minimum width=2cm, minimum height=1cm, text centered, draw=black, fill=green!30]
\tikzstyle{arrow} = [thick,->,>=stealth]

\begin{document}
\setcounter{page}{1}
\title[]{DIFFEOMORPHIC RECONSTRUCTION OF A 2D-SIMPLE
NON-PARAMETRIC MANIFOLD FROM A NOISY LEVEL SET DATA VIA SHAPE GRADIENTS}
\author[Shafeequdheen P]{Shafeequdheen P, Jyotiranjan Nayak, Vijayakrishna Rowthu}
\address{Department of Mathematics, SRM University-AP, Amaravati 522 240, India}
\email{shafeequdheen\_p@srmap.edu.in}

\begin{abstract}
A variational approach to the reconstruction of a shape (2D simple manifolds) as triangulated surface from given level set using shape gradients is presented. It involves an energy functional that depends on the local shape characteristics of the surface. Minimization of the energy through an iterative procedure using the gradient descent method yields a triangulated surface mesh which  matches the boundary of the object of interest and  this model ensures the smoothness of the boundary.
\end{abstract}

\maketitle

\section{Introduction}\label{Intro}

Shape reconstruction from discrete data plays a pivotal role in both computer graphics and medical image processing. In computer graphics, the creation of realistic 3D models is indispensable for applications such as video games and virtual reality. Converting scattered data points into smooth and continuous surfaces enhances visual quality and realism. Similarly, in medical image processing, including the interpretation of Magnetic Resonance Imaging (MRI) and Computed Tomography (CT) scans \cite{HAN2004997,Osechinskiy2012CorticalSR,781013}, precise shape reconstruction is crucial for accurately representing anatomical structures. The challenges inherent in shape reconstruction, such as insufficient data, noise in datasets, and surface complexity, have spurred numerous research endeavors. Two primary approaches to shape reconstruction are deformable and non-deformable models
\citep{han2003topology,montagnat2001review,bardinet1998parametric}. Among non-deformable techniques, the Marching Cubes Method (MCM) \cite{nielson2003marching} stands out for its efficiency in iso-surface extraction.

The marching cubes method is used to create a 3D surface from volumetric data that has been sampled on a grid. This algorithm works by processing each small cube, or voxel, in the grid separately. For each voxel, the algorithm examines the function values at its eight corners to determine how the surface passes through the voxel. There are 256 possible configurations for how the surface can intersect a voxel, and these configurations are precomputed and stored in a lookup table, allowing the algorithm to quickly identify the correct pattern. Linear interpolation is then applied along the edges of the voxel to accurately place the vertices of the surface. These vertices are connected to form triangles, creating a mesh that approximates the isosurface. The method is efficient, capable of handling complex shapes and widely utilized in various image processing software programs, MCM efficiently extracts surfaces from volumetric data. However, it does not inherently ensure the smoothness of the reconstructed shapes, posing a limitation in some applications.\\

DistMesh \cite{persson2004simple} is a versatile tool for generating triangulated mesh surfaces from level set. Developed by Persson and Strang, it uses a deformable model with iterative adjustment of nodes to achieve uniform distribution, accommodating complex geometries and boundary conditions. The procedure begins by defining a signed distance function, \( f(x,y) \), which characterizes the geometry of the domain. The Nodes $\{p_{0}\}$ are initialized within the domain, a Delaunay triangulation\citep{fortune2017voronoi,sloan1993fast,borouchaki1995fast} if fount and the algorithm iteratively adjusts their positions to achieve a uniform distribution. For each node \( p_0(x_{0},y_{0}) \), the closest point \( P(x,y) \) on the zero level set of \( f \) is found, ensuring that \( f(P) = 0 \) and that \( P - p_0 \) is parallel to the gradient \( \nabla f \) at \( P \). This is achieved by solving \( L(P) = 0 \) using a damped Newton's method, where \( L(P) \) is defined as:
\begin{equation}
L(P) = 
\begin{bmatrix}
f(x, y) \\
(x - x_0) f_y - (y - y_0) f_x
\end{bmatrix}
\end{equation}
The Jacobian of \( L \) is:
\begin{equation}
J(\mathbf{P}) = \frac{\partial L}{\partial \mathbf{P}} = 
\begin{bmatrix}
f_x & f_y + (x - x_0) f_{xy} - (y - y_0) f_{xx} \\
f_y & -f_x - (y - y_0) f_{xy} + (x - x_0) f_{yy}
\end{bmatrix}^T
\end{equation}
Iterate
\begin{equation}
p_{k+1} = p_k - \alpha J^{-1}(p_k) L(p_k)
\end{equation}
until the residual \( L(p_k) \) is small. 
%
The algorithm's flexibility allows for easy adaptation to various domains and boundary conditions, making it a powerful tool in computational geometry and finite element analysis.\\

The shape gradient approach, introduced by E. Debreuve et al. \cite{debreuve2007using}, is geared towards image segmentation through an energy functional formulation:

\begin{equation}\label{2D Energy}
E(\Gamma)=\int_{\Omega}\phi(\Gamma,x)dx+\int_{\Gamma}\varphi(s)ds
\end{equation}
In this expression, $\Omega$ denotes an open set in $\mathbb{R}^{2}$, $\Gamma$ represents the oriented boundary $\partial \Omega$ of $\Omega$, $s$  signifies the arc-length parametrization of $\Gamma$ and $f$ is the image to be segmented. The function $\phi$  defined by
\begin{equation}
\phi(\Gamma,x)=(f(x)-\mu(\Gamma))^{2}
\end{equation}
$\mu(\Gamma)$ is the average of value $f$ in $\Omega$, then $\phi$ is equal to zero on $\Omega$ if and only if  $f$ is constant on $\Omega$, and $\varphi$ is the descriptor of the object boundary. The shape gradient of \ref{2D Energy} is defined as the following ways;
\begin{equation}
dE(\Gamma;V)=\left. \int_{\Omega} \frac{\partial \phi(\Gamma(\tau), x)}{\partial \tau} \right|_{\tau=0}dx-\int_{\Gamma}\left(\phi(\Gamma,x)-\frac{\partial\varphi(\Gamma,s)}{\partial N}+ \varphi(s)\kappa(s)\right)N(s)\cdot V(s)ds
\end{equation}
Here, $V$ is a velocity field defined on $\Omega$, $N$ is the inward unit normal of $\Gamma$, and $\kappa$ is the curvature of $\Gamma$. Utilizing an active contour approach to minimize the energy \ref{2D Energy} involves iteratively deforming the initial polygon $\{\Gamma_{i}^{0}\}$ using the evolution equation:
\begin{equation}
\Gamma_i^{n+1} = \Gamma_i^{n} -\tau\  dE(\Gamma,V)V(\Gamma)
\end{equation}
The optimal value for $\tau$ can be computed as follows:
\begin{equation}
\tau_{min} = \arg\min_{\tau \geq 0} E(\Gamma^{n+1}(\tau))
\end{equation}
If $\tau_{min}$ is less than $\tau_{thresh}$, then the iteration is considered to have converged.

Accurately reconstructing the three-dimensional (3D) shape of highly intricate surfaces presents a significant challenge in the presence of noise. To address this challenge we introduce a novel approach: a topology-preserving parameter-free deformable model primarily designed for reconstructing simple two-dimensional (2D) manifolds. Our method also leverages the shape gradient of novel energy that depend on parameter-free surface. Moreover, the curvature of the surface appears naturally in the derived evolution equation. Which play important role in keeping the surface smooth. By integrating these components, this approach ensures not only enhanced the accuracy of surface but also resulted in smoother shape reconstructions for complex 3D shapes for noisy data. We rigorously evaluated the efficacy of our model using a series of phantoms constructed via  known mathematical functions and combinations. In the subsequent sections, we delve into the intricacies of our methodology, elucidating the formulation of the signed distance function (refer to section \ref{siged distance function})
 \subsection{Formation of Signed Distance Function (SDF)}\label{siged distance function}
The level set method, pioneered by S. Osher and J. A. Sethian \cite{zhao2000implicit,osher2004level,wang2003level}, stands as a robust technique for representing and tracking shape evolutions and object boundaries.
Consider a closed domain  $\Omega$  with boundary $\partial \Omega$. A distance function $d(x)$  defined as $d(x) = \{\min(|x - x_{I}|):x \in \partial \Omega,x_{I} \in \partial \Omega\}$, implying that  $d(x) = 0$ on the boundary.
A signed distance function, denoted by $\phi$, is an implicit function satisfying  $|\phi(x)| = d(x)$ for all $x$. Thus, $\phi(x) = d(x) = 0$ for all $x_{I} \in \partial \Omega$, $\phi(x) = -d(x)$ for all $x \in \partial \Omega^{-}$, and $\phi(x) = d(x)$ for all $x \in \partial \Omega^{+}$. Essentially, the zero-level set characterizes the closed boundary, while positive and negative levels represent the exterior and interior of the domain, respectively.

\begin{figure}[H]
\centering
\includegraphics[scale=0.5]{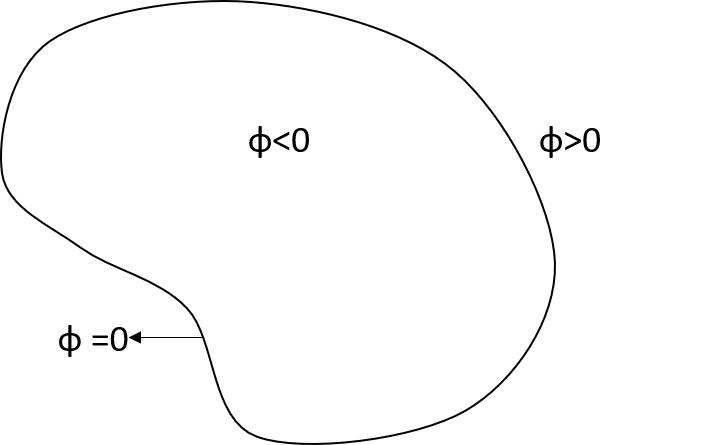}
\caption{Signed Distance Function $\phi$}\label{Phantoms}
\end{figure}

Set operations such as union, intersection, difference, and complement for signed distance functions can be defined as follows: If $\phi_{1}(x)$ and $\phi_{2}(x)$ are two distinct SDF, then $\phi(x) = \min(\phi_{1}(x),\ \phi_{2}(x))$ represents the SDF for the union of their interior regions. Similarly, $\phi(x) = \max(\phi_{1}(x),\ \phi_{2}(x))$ characterizes the intersection of the interior regions. The complement of the set defined by $\phi_{1}(x)$ has a SDF $\phi(x) = -\phi_{1}(x)$. Additionally,\ $\phi(x) = \max(\phi_{1}(x),\  -\phi_{2}(x))$ is the SDF for the region defined by subtracting the interior of $\phi_{2}(x)$ from the interior of $\phi_{1}(x)$.
Using these operations, various types of level sets were constructed for different phantoms for numerical experiments. 
\begin{enumerate}
\item \textbf{Unit sphere centered at the origin:} $\phi({x})=x_1^{2}+x_2^{2}+x_3^{2}-1$ \\ 
\item \textbf{Ellipsoid:} $\phi(x)=x_1^2+\dfrac{x_{2}^2}{2^2}+x_3^2-1$ \\ 
   \item \textbf{Two fused sphere:}
    Consider the signed distance functions of two spheres $\phi_{1}(x)$,\  $\phi_{2}(x)$ with radius $0.8$, centred at $(0,0,0.7)$ and $(-0.7,0,0)$ respectively. That is 
    $\phi_1({x})=x_1^{2}+x_2^{2}+(x_3-0.7)^{2}-0.8^2,\ \phi_2({x})=x_1^{2}+x_2^{2}+(x_3+0.7)^{2}-0.8^2$. $\phi(x)=min(\phi_{1}(x),\ \phi_{2}(x))$
    gives a signed distance function for the fused structure.\\ 
  \item \textbf{Cylinder:}
    Let $\phi_1({x})=x_3-1,\ \phi_2({x})=x_3+1$, and $\phi_3({x})=x_1^{2}+x_2^{2}-0.4^{2}$. Then $\phi(x)=max(\phi_1(x),\ \phi_2(x),\ \phi_3(x))$ represents the finite cylinder with radius $0.2$   and length $2$ .   
\end{enumerate}
The Table \ref{levelset_table} displays an unsigned distance function and level set of the phantom
\begin{table}[h!] 
    \centering
    \begin{tabular}{|c|c|c|}
         \hline
         Phantoms & Unsigned level set & Signed level set  \\
         \hline
         Sphere & \includegraphics[scale=0.4]{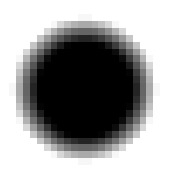} & \includegraphics[scale=0.4]{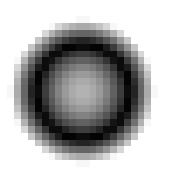}  \\
         \hline
         Ellipsoid & \includegraphics[scale=0.3]{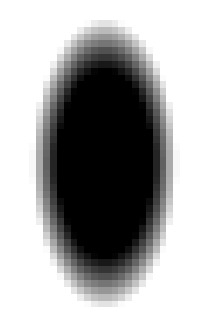} & \includegraphics[scale=0.3]{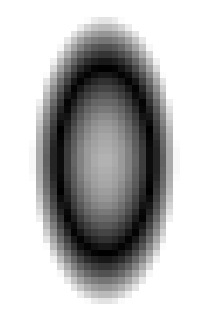}  \\
         \hline
         Two fused Sphere& \includegraphics[scale=0.4]{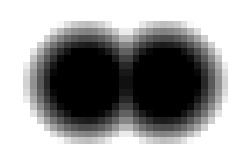} & \includegraphics[scale=0.4]{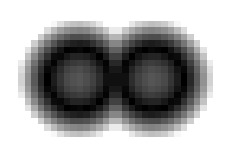}  \\
         \hline
          Cylinder & \includegraphics[scale=0.4]{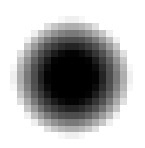} & \includegraphics[scale=0.4]{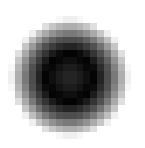}  \\
         \hline
    \end{tabular}
    \caption{Single slice image(along the z-axis) of unsigned distance and level set of phantoms.}
    \label{levelset_table}
\end{table}

\subsection{Mean Curvature Estimation on a Triangulated Surface} \label{Mean Curvature Estimation}

Curvature provides valuable information about the local geometry of a surface. It describes how much a curve deviates from being a straight line at a particular point, capturing the bending and twisting of the surface.

\begin{figure}[h]
  \centering
  \includegraphics[scale=0.2]{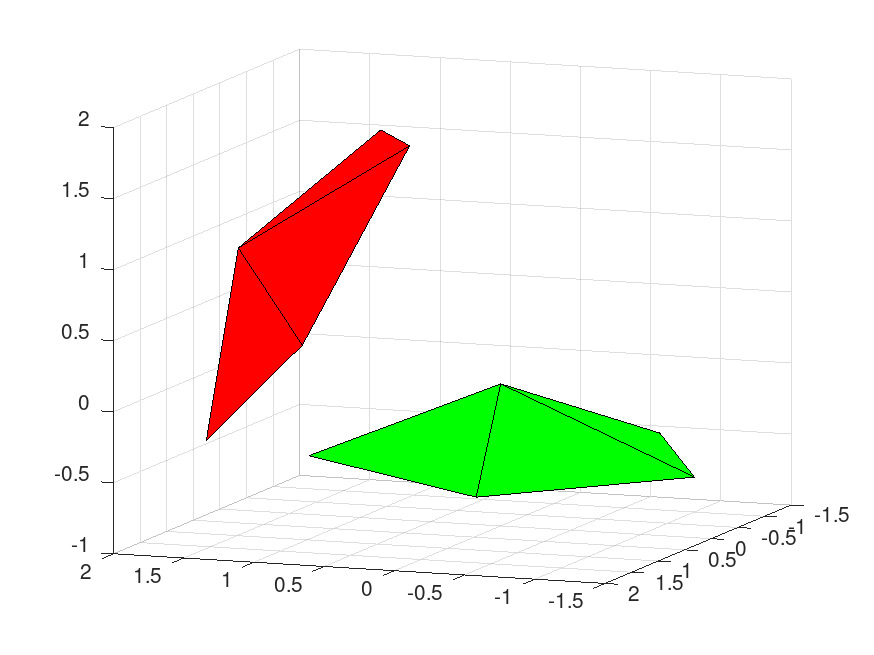}
 \caption{Rotation of patch: Red patch (before rotation) and green patch (after rotation).}
  \label{Rotation patch}
\end{figure}

To find the mean curvature $H$ at the vertex $v$ on a triangulated mesh surface \cite{dong2005curvatures,garimella2003curvature,hamann1993curvature}, approximate each local patch around vertex $v$ with a paraboloid \cite{goldfeather2004novel} of the form:
\begin{equation}\label{paraboloid}
z(x, y) = \frac{A}{2}x^{2} + Bxy + \frac{C}{2}y^{2}
\end{equation}
Such that, the vertex $v$ and all neighboring vertices of $v$ lie on the surface of the paraboloid \eqref{paraboloid}. Since curvature is invariant under rotation and translation, each neighboring patch of vertex $v$ can be uniquely shifted as follows:\\
Let $v$ be the center vertex of a triangular patch, and let $v_{i}$ be the adjacent vertices to $v$. Shift the vertices
$v$ to the origin, and align the normal at $v$ of the patch along the vector $\hat{k}$. The local coordinates of the neighboring vertices $v_{i}$ can be found using the rotation matrix $R$ derived from Rodrigues' rotation formula:
\begin{equation}\label{Rodrigues' rotation formula}
R = I + (\sin\theta)K + (1-\cos\theta)K^{2}
\end{equation}
Here, $\theta$ represents the angle between $\hat{k}$ and the weighted average normal $\hat{n}$ around vertex $v$. $\hat{\rho}$ defines the axis of rotation and is given by:
A shape-dependent variational approach to reconstruct a simple 2D manifold from level set discrete data is presented. The numerical experiments reveal that, in each iteration, the tangential gradient term plays an important role in the smooth deformation of the surface, resulting in diffeomorphic evolution. The implications of this approach extend to the accurate representation of complex surfaces, such as the cortex of the human brain, with potential applications in medical image processing
\begin{equation}
 \hat{\rho}= \frac{\hat{n} \times \hat{k}}{|\hat{n} \times \hat{k}|}=[\rho_{x},\  \rho_{y},\  \rho_{z}]
\end{equation}
and
\begin{equation}
K = \begin{bmatrix}
0 & -\rho_{z} & \rho_{y}\\
\rho_{z} & 0 & -\rho_{x}\\
-\rho_{y} & \rho_{x} & 0
\end{bmatrix}
\end{equation}
The new coordinates are obtained by multiplying the rotation matrix $R$ with corresponding vertices of patch. In Figure \ref{Rotation patch}, the red-colored patch represents the state before rotation, and the green-colored patch represents the state after rotation.

Fit a paraboloid surface around every vertex of the local patch lying on the surface. Using the least squares method, determine the values of unknowns $A$, $B$, and $C$. The Weingarten matrix for the paraboloid (\ref{paraboloid}) is:
\begin{equation}
W = \begin{bmatrix}
-A & B\\
B & -C
\end{bmatrix}
\end{equation}
Therefore, the Mean curvature $(H)$ and Gaussian curvature $(G)$ can be defined as:
\begin{equation}\label{H}
H = \frac{1}{2} \text{trace}(W), \quad G = \text{det}(W)
\end{equation}
Find the Weingarten matrix for each vertex $v$, curvature at the vertex  by $-0.5(A + C)$.\\
%
%
\section{Shape Derivative} \label{shape Derivative}
The shape derivative or gradient \cite{debreuve2007using,aubert2003image} measures how the energy functional changes with respect to changes in the shape of the domain. It tells us how sensitive the shape functional is to small variations or deformations in the shape.\\
Consider the domain $\Omega\subset \mathbb{R}^{3}$ and $\Gamma$ represents the boundary of the domain and $T_{t}$ be the family transformation  defined by  
\begin{equation}
    T_{t}:\mathbb{R}^3 \longrightarrow \mathbb{R}, \quad t\in [0,\epsilon )
\end{equation}
$T_{t}$ maps the point $X\in \Omega$ onto $x\in \Omega_{t}$, where $T_{t}(\Omega)=\Omega_{t}$.
Let $V(t,x)$ be the velocity at the point $x(t)$, defined by 
\begin{equation}
    V(t,x)=\frac{\partial x}{\partial t}(t,T^{-1}_{t}(x))
\end{equation}
Define the shape derivative \cite{Sokolowski1992,chicco2017shape,dogan2007discrete} of a functional $J$ in the direction of the vector field $V$ as follows:
\begin{equation}\label{shape derivative}
    dJ(\Omega;V)=\lim_{t\rightarrow 0} \frac{1}{t}(J(\Omega_{t})-J(\Omega)).
\end{equation}
Let $\Gamma$ be of class $C^{2}$ and $\phi\in W^{2,1}(\mathbb{R}^{3})$. Consider the functional 
\begin{equation}
    J(\Gamma)=\int_{\Gamma}\phi\  d\Gamma
\end{equation}
is shape differentiable for perturbation vector field $V\in C_{0}^{1}(\mathbb{R}^{2};\mathbb{R}^{2})$ with shape differentiable functional
\begin{equation} \label{dJ1}
    dJ(\Gamma;V)=\int_{\Gamma}(\nabla \phi \cdot V +\phi\  \text{div}_{\Gamma}V)\ d\Gamma.
\end{equation}
Similarly, if $\Gamma$ be of class $C^{2}$ and $\phi\in W^{2,2}(\mathbb{R}^{3})$. Then the functional \cite{Sokolowski1992},
\begin{equation} 
J(\Omega)=\int_{\Gamma}(\nabla \phi \cdot \hat{n})^{2} d\Gamma
\end{equation}
is shape differentiable for perturbation vector field $V\in C_{0}^{1}(\mathbb{R}^{2};\mathbb{R}^{2})$ with
\begin{equation} \label{dJ2}
dJ(\Gamma;V)=\int_{\Gamma} \left\{2 \frac{\partial \phi}{\partial \hat{n}} [D^{2}\phi] \hat{n} \cdot \hat{n}+H\left|\frac{\partial \phi}{\partial \hat{n}}\right|^{2}+2 \text{div}_{\Gamma}\left(\frac{\partial \phi}{\partial \hat{n}}\nabla_{\Gamma}\phi\right)\right\}V\cdot \hat{n} \ d\Gamma
\end{equation}
\section{Gradient Descent Method}
The shape gradient method is a deformable technique used to reconstruct a 2D simple manifold in the form of triangulated surface from a given level set. A triangulated sphere serves as the initial approximated to target the  surface, which is then deformed against the direction of the shape gradient, denoted as $-\overrightarrow{dE}$, computed via gradient descent method.
That is,
\begin{equation}
dE(\Gamma;V)=\langle {\overrightarrow{dE(\Gamma)},V} \rangle
\end{equation} 
Let $\Gamma_{0}=\Gamma(V_{0},T)$ represent the initial surface, where $V_{0}$ encompasses all vertices of the triangles on the initial triangulated mesh surface $\Gamma_{0}$, and $T$ consists of all triangular elements. In each iteration, the vertex set $V_{0}$ is updated utilizing the gradient descent minimization method, defined as:
\begin{equation}
V_{n+1}=V_{n}-\overrightarrow{dE}(\Gamma)
\end{equation}

This iterative process terminates when $V_{n+1}\approx V_{n}$, achieving a minimum energy state.

\subsection{Mathematical Model}
Let $\Omega\subset \mathbb{R}^{3}$ be the domain, $\Gamma$ denotes its boundary, and $\phi$ represent the signed distance function associated with an object within $\Omega$. The energy functional is defined as:
\begin{equation}
E(\Gamma)=\int_{\Gamma}(\alpha\ \phi^{2}(\Gamma)+\beta\ |\nabla_{\Gamma}\phi|^{2})d\Gamma \label{Model}
\end{equation}
Here, $\alpha$, $\beta$  are constants and $\phi$ is fixed level set. The term $\phi^{2}(\Gamma)$ is to ensure the surface $\Gamma$ stays closer to zero level set, the term $|\nabla_{\Gamma}\phi|^{2}$ ensures the surface $\Gamma$ assumes  a constant level, may also be a nonzero. Considering $|\nabla_{\Gamma}\phi|^{2}=|\nabla\phi|^2-(\nabla\phi \cdot \hat{n})^{2}$, where $\hat{n}$ is the unit normal vector to the surface $\Gamma$, the energy functional \ref{Model} simplifies to:
\begin{equation}
E(\Gamma)=\int_{\Gamma}( \alpha \ \phi^{2}(\Gamma)+\beta \ |\nabla\phi|^2-\beta \ (\nabla\phi \cdot \hat{n})^{2})d\Gamma
\end{equation}
Now the objective is to minimize $E(\Gamma)$ using shape gradient descent method. Notably, since $\phi$ is zero on the object's surface, $E(\Gamma)$ is ideally minimum when $\phi=0$ and, level set gradient and normal of the surface $(\Gamma)$ are parallel. The first term in \ref{Model} aids in locating the boundary, while the second term promotes the smoothness of the deformable surface $\Gamma$. Utilizing \ref{dJ1} and \ref{dJ2}, the energy functional $E(\Gamma)$ is shape-differentiable for any vector field $V$. Considering $V$ as the unit normal vector $\hat{n}$ around the surface $\Gamma$, the shape gradient of \ref{Model} is given by:


\begin{equation} \label{dJ}
\begin{aligned}
    dE(\Gamma;\hat{n}) = & \int_{\Gamma} \left\{ \left( \nabla \phi^{2} + \nabla |\nabla \phi|^{2} \right) \cdot \hat{n} + H\left( \phi^{2} + |\nabla\phi|^2 - \left|\frac{\partial\phi}{\partial\hat{n}}\right|^2 \right) \right. \\
    & \left. - 2\frac{\partial\phi}{\partial\hat{n}} \left( \nabla\phi \cdot \left( ^{*}[D\hat{n}]\hat{n}+[D\hat{n}]\hat{n} \right) - [D^{2}\phi] \hat{n}\cdot \hat{n} \right) \right\} d\Gamma
\end{aligned}
\end{equation}
Here, $H$ denotes the Mean curvature, $[D\hat{n}]$ signifies the gradient (tensor) of the vector $\hat{n}$ , and $^{*}[D\hat{n}]$ denotes the conjugate transpose of $[D\hat{n}]$. Note that, $[D\hat{n}]\hat{n}=0$ for the unit normal $\hat{n}$ on the paraboloid \ref{paraboloid}. Hence $dE$ is the Riesz representative  vector element as follows.
\begin{equation} \label{dJ_finel}
\begin{aligned}
    \langle \overrightarrow{dE(\Gamma)} , \hat{n} \rangle = & \int_{\Gamma} \left\{ \left( \nabla \phi^{2} + \nabla |\nabla \phi|^{2} \right) \cdot \hat{n} + H\left( \phi^{2} + |\nabla\phi|^2 - \left|\frac{\partial\phi}{\partial\hat{n}}\right|^2 \right) \right. \\
    & \left. - 2\frac{\partial\phi}{\partial\hat{n}} \left(  [D^{2}\phi] \hat{n}\cdot \hat{n} \right) \right\}\hat{n}\cdot \hat{n}\ d\Gamma
\end{aligned}
\end{equation}

The subsequent steps outline the primary procedures of the numerical algorithms.

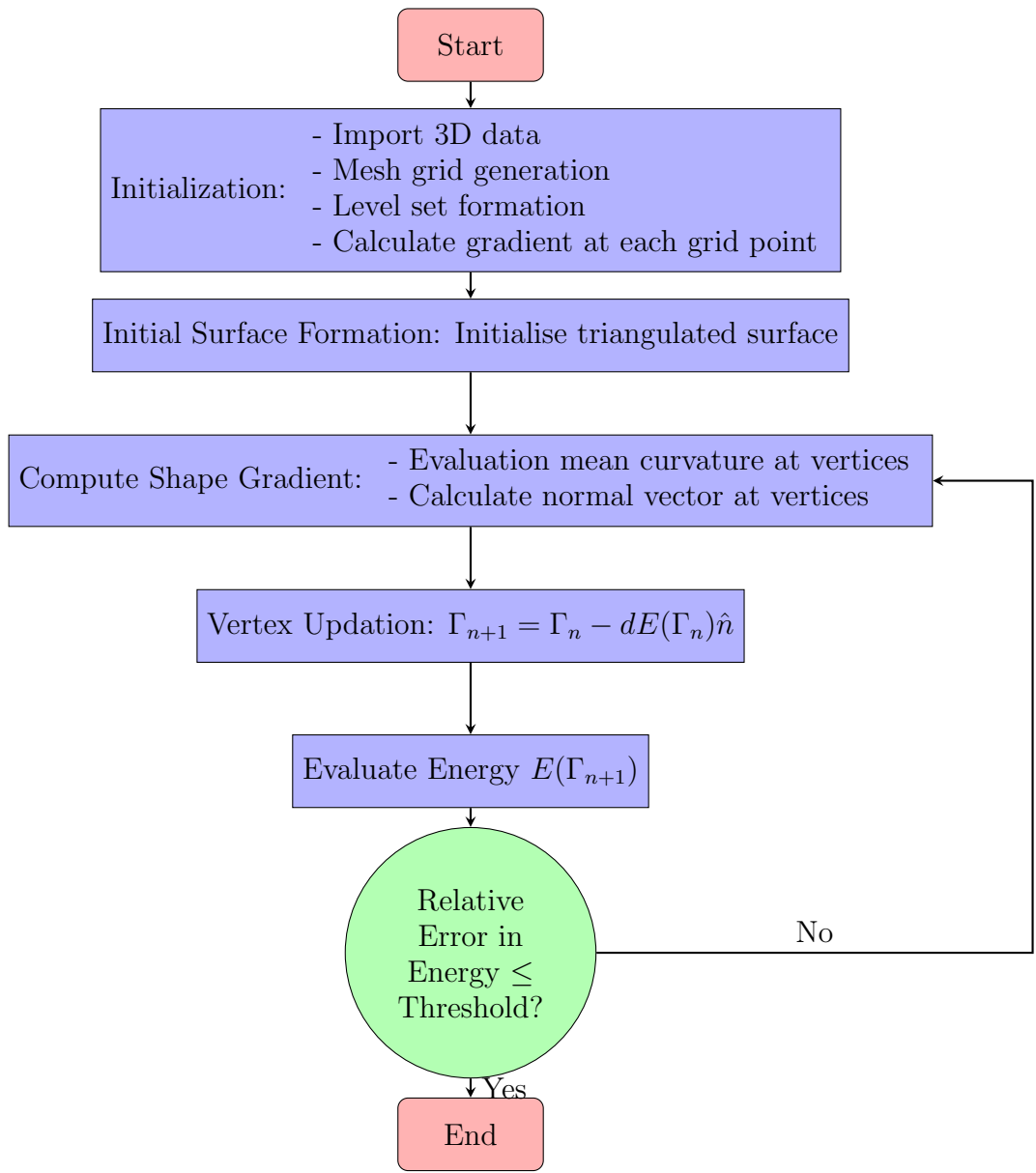
\begin{figure}
\centering
\begin{tikzpicture}[node distance=2cm]
\node (start) [startstop] {Start};
\node (init) [process, below of=start] {Initialization: 
\begin{tabular}{l}
    - Import 3D data\\
    - Mesh grid generation\\
    - Level set formation\\
    - Calculate gradient at each grid point
\end{tabular}};
\node (triangulate) [process, below of=init] {Initial Surface Formation: Initialise triangulated surface}; 

\node (gradient) [process, below of=triangulate] {Compute Shape Gradient:
\begin{tabular}{l}
    - Evaluation mean curvature at vertices\\
    - Calculate normal vector at vertices
\end{tabular}};
\node (vertex) [process, below of=gradient] {Vertex Updation: $\Gamma_{n+1}=\Gamma_n-dE(\Gamma_{n})\hat{n}$};
\node (energy) [process, below of=vertex] {Evaluate Energy $E(\Gamma_{n+1})$};
\node (check) [decision, below of=energy, yshift=-0.5cm, text width=2.5cm] {Relative Error in Energy $\leq$ Threshold?};
\node (end) [startstop, below of=check, yshift=-0.5cm] {End};

\draw [arrow] (start) -- (init);
\draw [arrow] (init) -- (triangulate);
\draw [arrow] (triangulate) -- (gradient);
\draw [arrow] (gradient) -- (vertex);
\draw [arrow] (vertex) -- (energy);
\draw [arrow] (energy) -- (check);
\draw [arrow] (check.east) -- ++(6,0) node[midway, above] {No} |- (gradient.east);
\draw [arrow] (check) -- node[midway, right] {Yes} (end);

\end{tikzpicture}
\caption{Flowchart for Surface Evolution}
\label{fig:surface-evolution-flowchart}
\end{figure}

\begin{enumerate}
    \item {Initialization:} Define a meshgrid and determine the level set values of a chosen phantom shape at the grid points. Establish the initial surface $\Gamma_{0}$ as a triangulated Sphere with a radius of two, serving as the baseline for the manifold reconstruction.
    
    \item {Level Set Initialisation:} Utilise trilinear interpolation to compute the level set values at each vertex of the triangulated sphere. This step lays the groundwork for subsequent computations.
    
    \item {Iterative loop for surface evaluation:} For finding the shape Riesz representative element $dE$ for each vertex, more number of interpolation are required in direction.
    \begin{enumerate}
        \item {Computation of normals at vertices}: The normal at a vertex is computed as the weighted average of normals of neighborhood elements.
        \item {Mean curvature computation}: Use the procedure \ref{H}.
        \item {Shape gradient computation:} Use \ref{dJ_finel} to execute the shape gradient as the Riesz representative element $dE$.
        \end{enumerate}

     \item {Vertex Update:} Update the vertices of the initial Sphere based on the computed shape gradient term. The updated vertices yield a new surface $\Gamma_{1}$, determined by the equation $V_{1} = V_{0} - \overrightarrow{dE}(\Gamma)$.
    
    \item {Energy assessment:} Evaluate the energy functional resulting from the updated surface $\Gamma_{1}$. 
    
    \item {Stopping criteria:} If the energy falls below a predefined threshold, indicating convergence, the iteration halts. However, if the energy remains significant, suggesting further refinement is required, proceed to the next iteration. Repeat the iterative loop until convergence.
\end{enumerate}

\begin{algorithm}
\caption{Surface Evolution Using Level Set and Shape Gradient}
\begin{algorithmic}[1]
\Require 3D data array, parameters $\alpha, \beta$, step size $\delta t$
\Ensure Evolved triangulated surface $V$
\State Initialize $\alpha$, $\beta$
\State Import 3D data and preprocess
\State  \quad Generate mesh grid for level set function $\phi$
\State \quad Evaluate $\phi$ at grid points and reshape for 3D representation
\State  \quad Compute signed distance function
\State Initialize triangulated surface $\Gamma=(V, E)$
\For{$s = 1$ to $100$} \Comment{Surface evolution loop}
    \State Compute mesh vertex normals $\hat{n}$
    \State Mean curvature $H$
    \State Calculate shape gradient $dJ$ 
    \State Update vertex positions:
    \State \quad $\Gamma_{n+1} = \Gamma_{n} - \delta t \cdot dJ$
    \State Evaluate energy $E(\Gamma)$
    \If{$E(\Gamma) < \text{threshold}$}
        \State \textbf{break}
    \EndIf
\EndFor
\end{algorithmic}
\end{algorithm}

\section{Numerical Results}
In this section, we present numerical results for several phantoms (2D Simple manifolds). Computation trials were done using GNU/Octave programming Software (similar to \textsc{Matlab}\textsuperscript{\textregistered}) on the Ubuntu 22.04 OS, Intel Core i7-12750 processor, 3.6 GHz with 20 cores and 64GB of memory. To expedite  this process , we implemented parallel computation in the algorithm using parallel package of GNU/Octave \cite{azzini2018dragonfly}. For the numerical experiments, we employed the signed distance functions of various phantoms as described in Section \ref{siged distance function}. All the experiments commenced with the same initial surface, $\Gamma_{0}$, which was a triangulated sphere with a radius of 2, comprising 1896 vertices and 3788 triangles. 

In each iteration, the initial surface $\Gamma_{0}$ undergoes deformation towards the ideal surface, and the energy graph depicts the smoothness behavior during the surface deformation in each iteration. The tangential gradient term in the mathematical model exhibits high sensitivity, ensuring not only the smoothness of the surface but also accelerating the deformation towards the ideal surface of the phantom. We have observed that this mathematical model is sufficiently effective in reconstructing the simple 2D manifold from SDF.\\
The following Table illustrates the reconstruction of several phantoms: Sphere, Ellipsoid, fused Sphere, and Cylinder. The second column of each table shows the deformation of the initial surface $\Gamma_{0}$ (a sphere with a radius of 2) into the ideal Sphere using the energy functional without the tangential gradient term, while the third column represents the deformation of the initial surface using the energy function \ref{Model} with stability parameters $\alpha=5$ and $\beta=1$. Additionally, the fourth column represents the deformation with noisy data for each phantom. Furthermore, from the energy graph, it is evident that there is a smooth deformation of the ideal surface from the initial surface.\\
The Table \ref{tab:Sphere_deformation} presents the reconstruction of a unit Sphere from level sets. We performed 100 iterations, both excluding and including the tangential component in our model. The initial surface $\Gamma_{0}$ is a Sphere with a radius of 2. It is evident that with each iteration, the initial surface deforms towards the ideal surface, and the color distribution in each triangle is based on curvature, where red represents high curvature and blue represents low curvature. By the 100th iteration, the curvature of the Sphere is nearly $1$. The fourth column represents the deformation of the unit Sphere with noisy data, demonstrating the effectiveness of our model even in the presence of noise. Similarly the Table \ref{tab:Ellipsoid}, \ref{tab:Fused Sphere}, \ref{tab:Cylinder} show the reconstruction of Ellipsoid, Fused Sphere and Cylinder from level set, and Tables \ref{tab:marching_cube} represents the reconstruction of the same phantoms using the Marching Cube Method (MCM) as discussed in section \ref{Intro}.

\begin{table}
    \centering
    \resizebox{\linewidth}{!}{%
    \begin{tabular}{|c|c|c|c|}
    	\hline
     		Iteration  & $\alpha=5,\ \beta=0$ & $\alpha=5,\ \beta=1$& $\alpha=5,\ \beta=1$, SNR$=44.5$ dB  \\
        \hline
         	Initial surface $\Gamma_{0}$ & \includegraphics[scale=0.5]{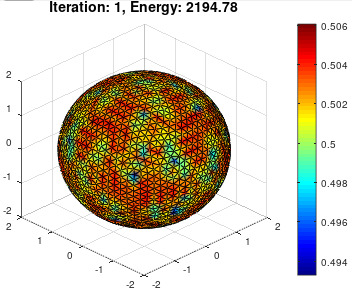} & \includegraphics[scale=0.5]{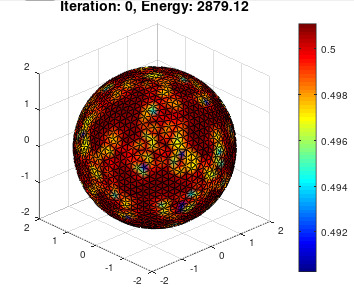} & \includegraphics[scale=0.5]{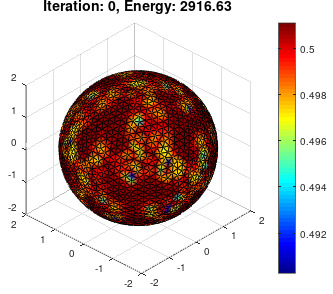}\\
         \hline
         $1^{st}$ Iteration & \includegraphics[scale=0.5]{pics/sphere_1_b=0.png} & \includegraphics[scale=0.5]{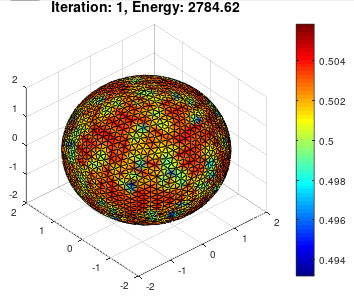}   & \includegraphics[scale=0.5]{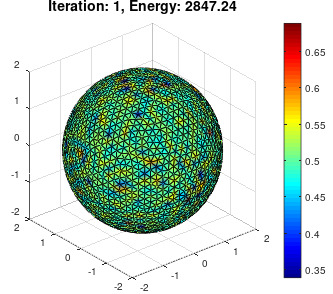}\\
         \hline
        $50^{th}$th Iteration & \includegraphics[scale=0.5]{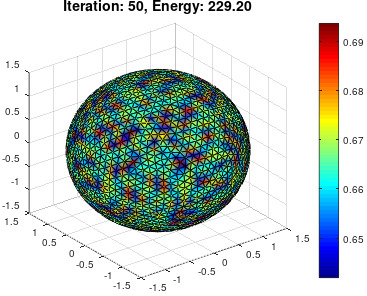} & \includegraphics[scale=0.5]{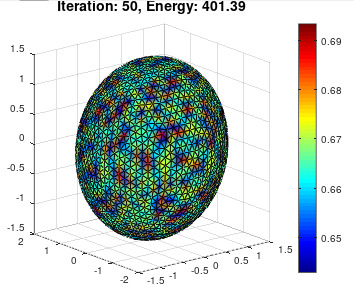}  & \includegraphics[scale=0.5]{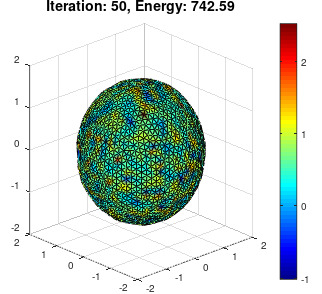}\\
         \hline
         $100^{th}$ Iteration & \includegraphics[scale=0.5]{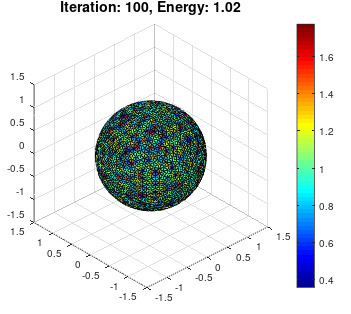} & \includegraphics[scale=0.5]{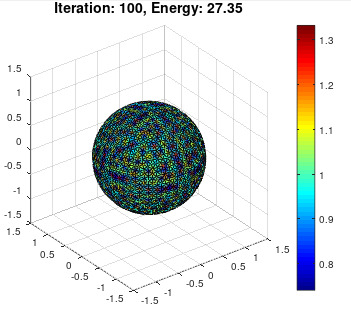}   & \includegraphics[scale=0.5]{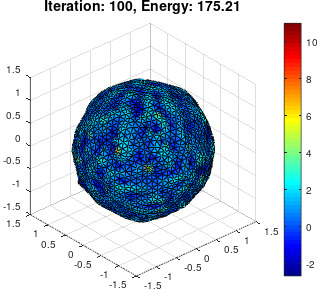}\\
         \hline
               Energy  & \includegraphics[scale=0.5]{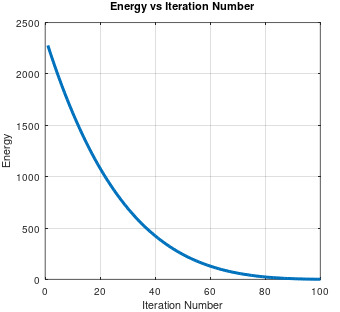} & \includegraphics[scale=0.5]{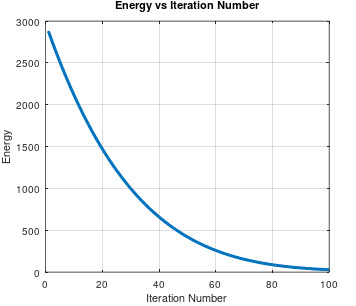} & \includegraphics[scale=0.5]{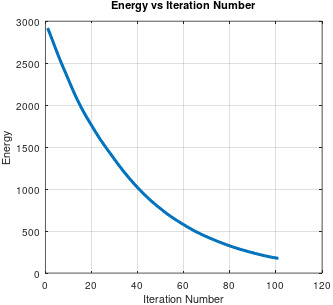} \\
         \hline
    \end{tabular}}
\caption{Reconstruction of the triangulated surface of a unit Sphere from level sets using shape gradient without tangential term, with tangential term, and with Gaussian noisy data.}
    \label{tab:Sphere_deformation}
\end{table}

\begin{table}
    \centering
    \resizebox{\linewidth}{!}{%
    \begin{tabular}{|c|c|c|c|}
    	\hline
     		Iteration  & $\alpha=5,\ \beta=0$ & $\alpha=5,\ \beta=1$ &  $\alpha=5,\ \beta=1$, SNR $=42$ dB\\
        \hline
         	Initial surface $\Gamma_{0}$ & \includegraphics[scale=0.5]{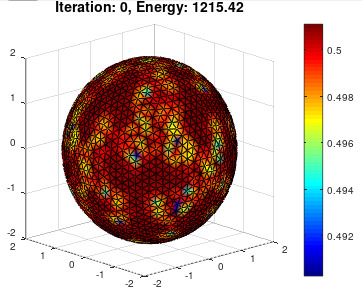} & \includegraphics[scale=0.5]{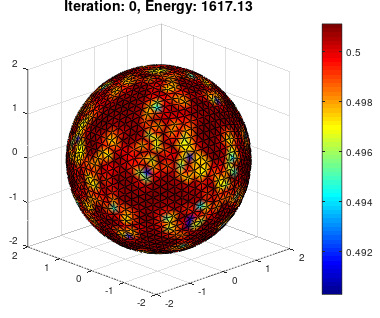} & \includegraphics[scale=0.5]{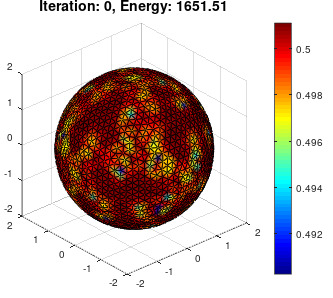}  \\
         \hline
         $1^{st}$ Iteration & \includegraphics[scale=0.5]{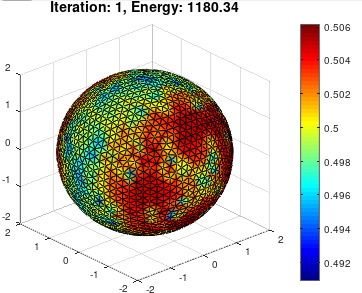} & \includegraphics[scale=0.5]{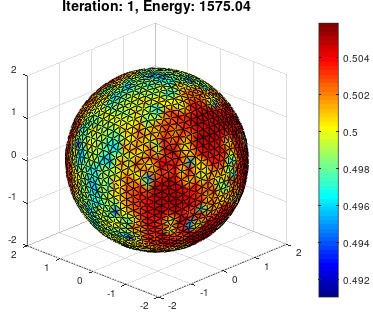} & \includegraphics[scale=0.5]{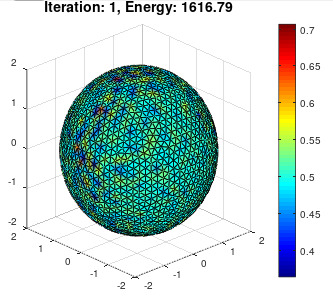}  \\
         \hline
        $50^{th}$ Iteration & \includegraphics[scale=0.5]{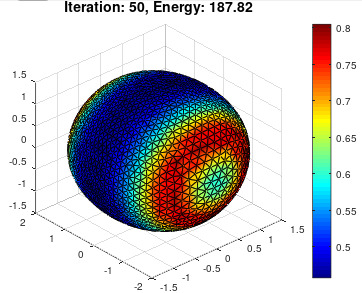} & \includegraphics[scale=0.5]{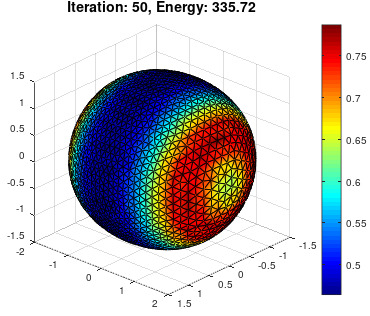} & \includegraphics[scale=0.4]{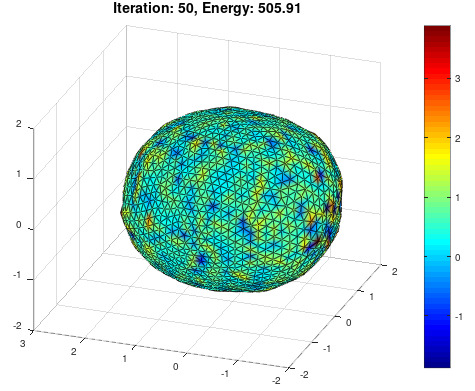}  \\
         \hline
         $100^{th}$ Iteration & \includegraphics[scale=0.5]{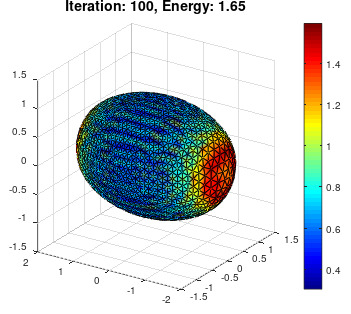} & \includegraphics[scale=0.5]{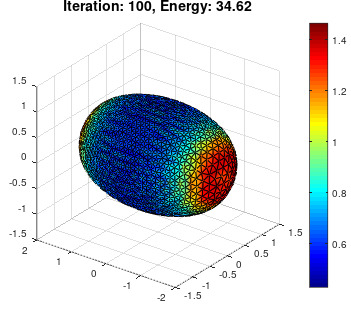} & \includegraphics[scale=0.4]{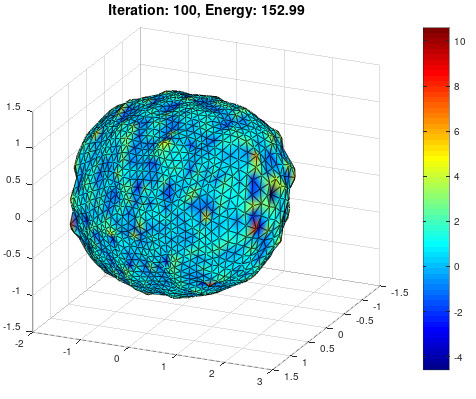}  \\
         \hline
               Energy  & \includegraphics[scale=0.5]{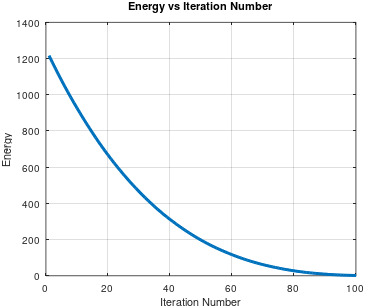} & \includegraphics[scale=0.5]{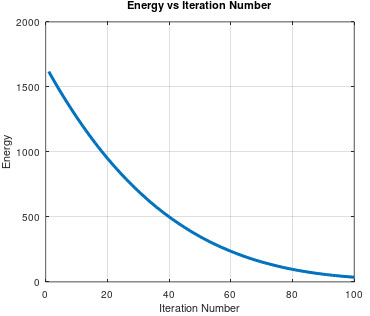} & \includegraphics[scale=0.4]{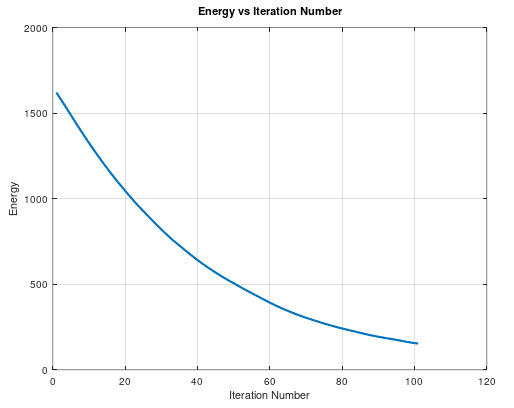}  \\
         \hline
    \end{tabular}}
\caption{Reconstruction of the triangulated surface of an Ellipsoid from level sets using shape gradient without tangential term, with tangential term, and with Gaussian noisy data.}
    \label{tab:Ellipsoid}
\end{table}

\begin{table}
    \centering
    \resizebox{\linewidth}{!}{%
    \begin{tabular}{|c|c|c|c|}
    	\hline
     		Iteration  & $\alpha=5,\ \beta=0$ & $\alpha=5,\ \beta=1$ &  $\alpha=5,\ \beta=1$, SNR $=43.2$ dB \\
        \hline
         	Initial surface $\Gamma_{0}$ & \includegraphics[scale=0.5]{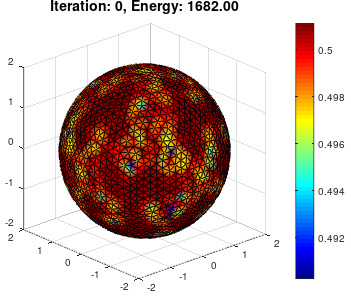} & \includegraphics[scale=0.5]{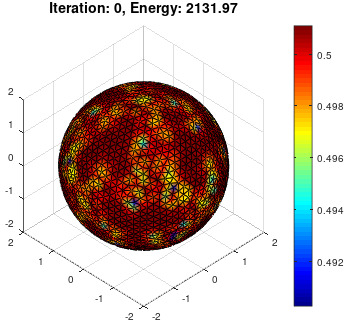} & \includegraphics[scale=0.5]{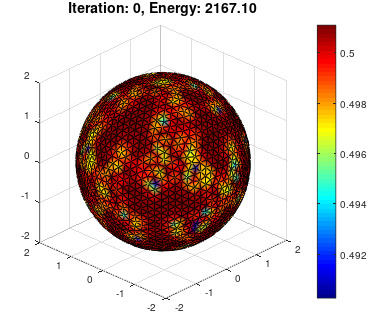}\\
         \hline
         $1^{st}$ Iteration & \includegraphics[scale=0.5]{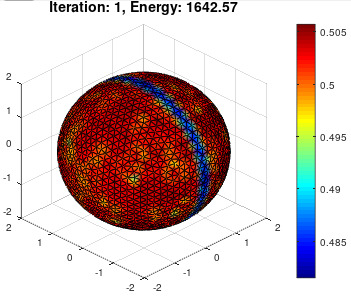} & \includegraphics[scale=0.5]{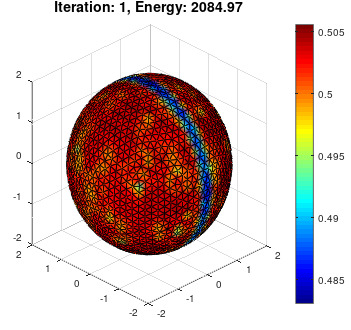}   & \includegraphics[scale=0.5]{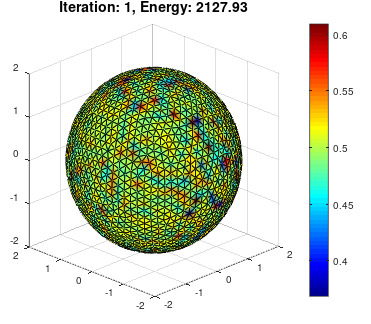}\\
         \hline
        $75^{th}$ Iteration & \includegraphics[scale=0.5]{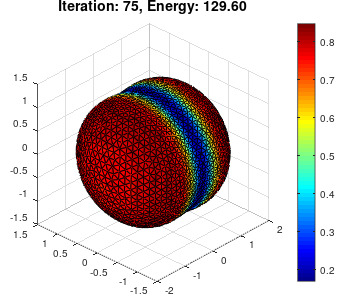} & \includegraphics[scale=0.5]{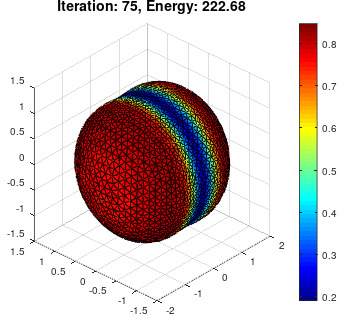}  & \includegraphics[scale=0.5]{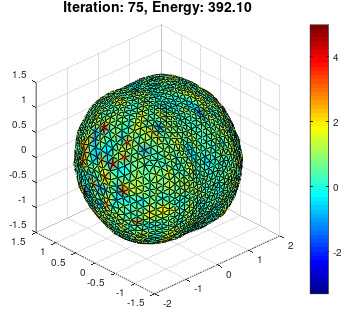} \\
         \hline
         $150^{th}$ Iteration & \includegraphics[scale=0.5]{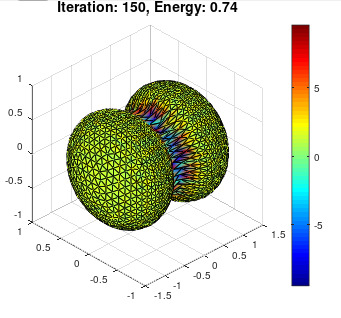} & \includegraphics[scale=0.5]{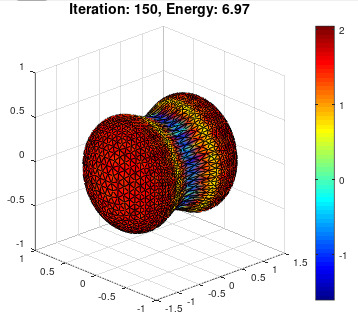}  & \includegraphics[scale=0.4]{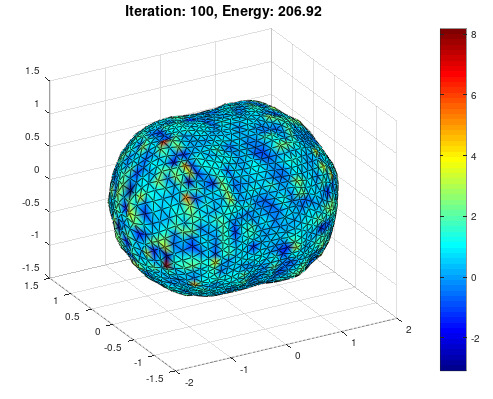}  \\
         \hline
               Energy  & \includegraphics[scale=0.5]{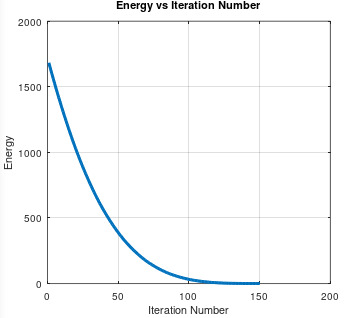} & \includegraphics[scale=0.5]{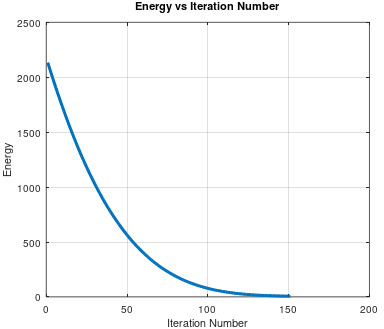} & \includegraphics[scale=0.39]{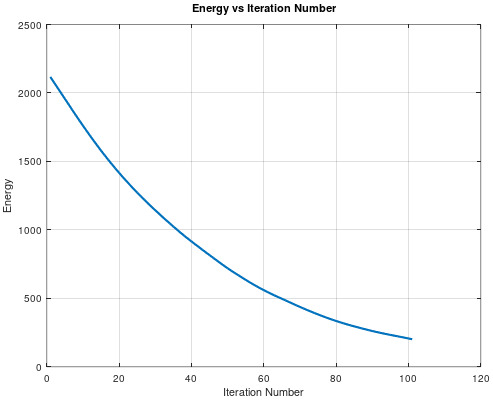}  \\
         \hline
    \end{tabular}}
\caption{Reconstruction of the triangulated surface of a Fused Sphere from level sets using shape gradient without tangential term, with tangential term, and with Gaussian noisy data.}
    \label{tab:Fused Sphere}
\end{table}

\begin{table}
    \centering
    \resizebox{\linewidth}{!}{%
    \begin{tabular}{|c|c|c|c|}
    	\hline
     		Iteration  & $\alpha=5,\ \beta=0$ & $\alpha=5,\ \beta=1$ &  $\alpha=5,\ \beta=1$, SNR $=42$ dB\\
        \hline
         	Initial surface $\Gamma_{0}$ & \includegraphics[scale=0.5]{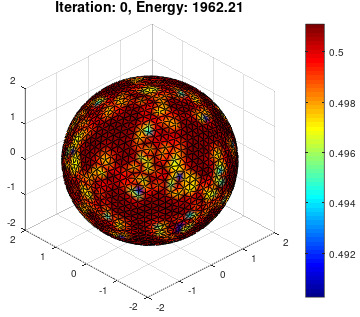} & \includegraphics[scale=0.5]{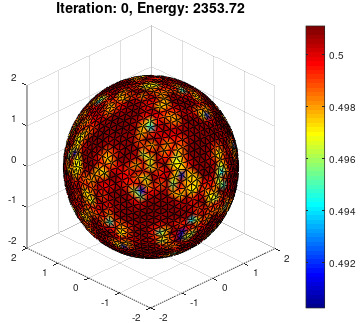} & \includegraphics[scale=0.5]{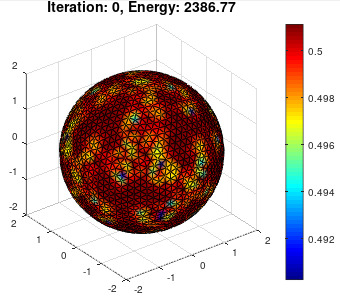}{a} \\
         \hline
         $1^{st}$ Iteration & \includegraphics[scale=0.5]{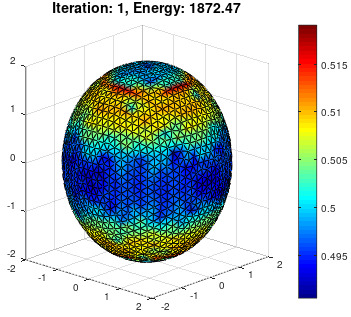} & \includegraphics[scale=0.5]{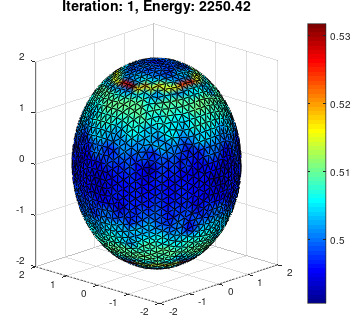} & \includegraphics[scale=0.5]{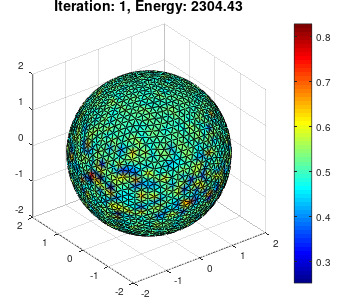}  \\
         \hline
        $50^{th}$ Iteration & \includegraphics[scale=0.5]{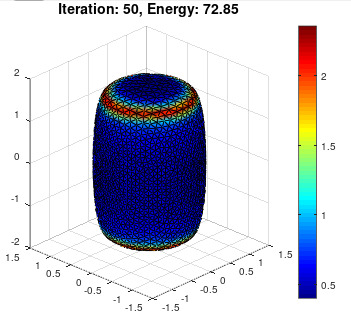} & \includegraphics[scale=0.5]{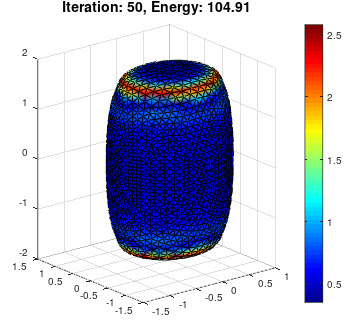} & \includegraphics[scale=0.4]{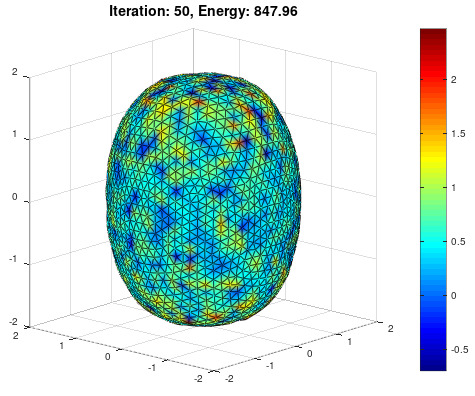}  \\
         \hline
         $100^{th}$ Iteration & \includegraphics[scale=0.5]{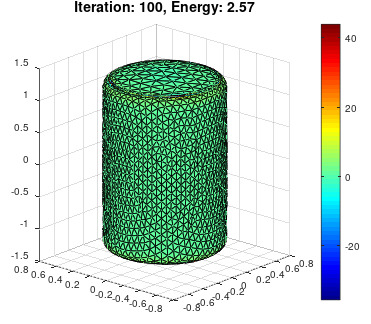} & \includegraphics[scale=0.5]{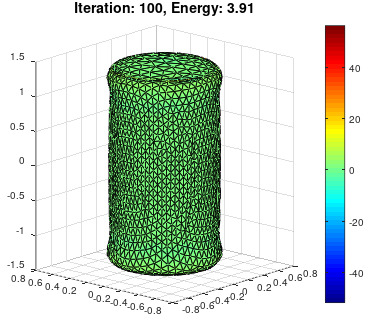} & \includegraphics[scale=0.4]{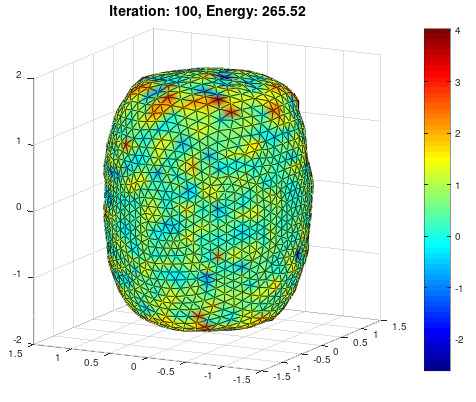}   \\
         \hline
               Energy  & \includegraphics[scale=0.5]{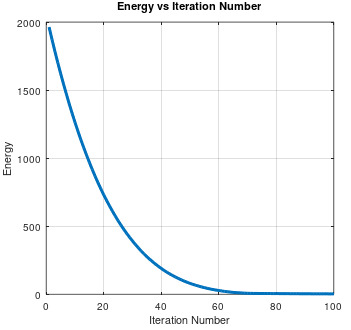} & \includegraphics[scale=0.5]{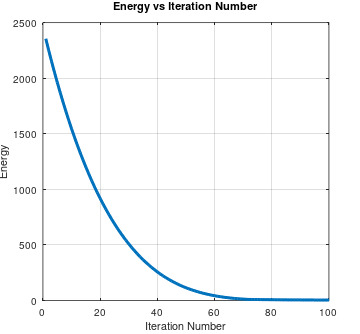} & \includegraphics[scale=0.4]{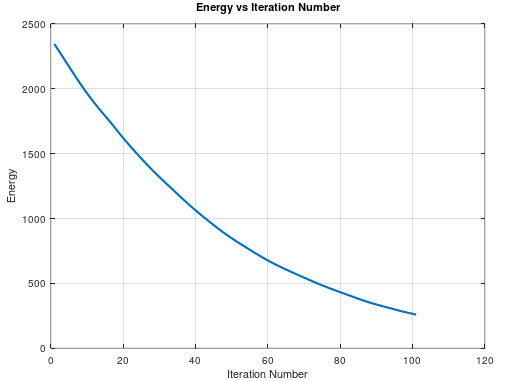}   \\
         \hline
    \end{tabular}}
\caption{Reconstruction of  triangulated surface of Cylinder from level sets using shape gradient without tangential term, with tangential term, and with Gaussian noisy data.}
    \label{tab:Cylinder}
\end{table}

\begin{table}
    \centering
    \resizebox{\linewidth}{!}{%
    \begin{tabular}{|c|c|c|c|}
    	\hline
     		Phantoms  & $\alpha=5,\ \beta=1$ & Energy\\
        \hline
         	Sphere (SNR $44.5$ dB) & \includegraphics[scale=0.5]{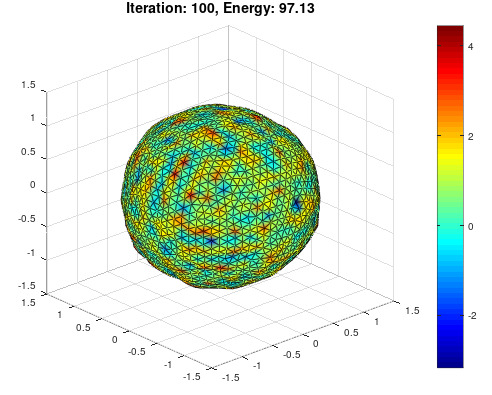} & \includegraphics[scale=0.5]{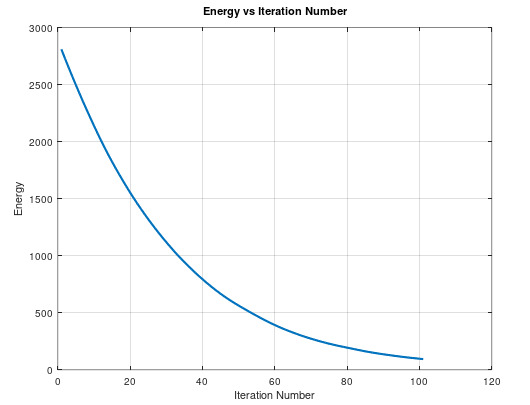}  \\
         \hline
         Ellipsoid (SNR $42.5$ dB) & \includegraphics[scale=0.5]{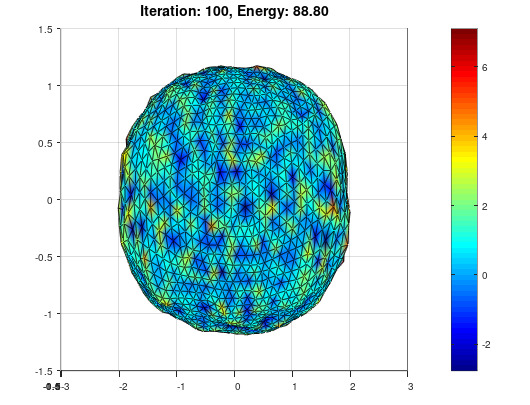} & \includegraphics[scale=0.5]{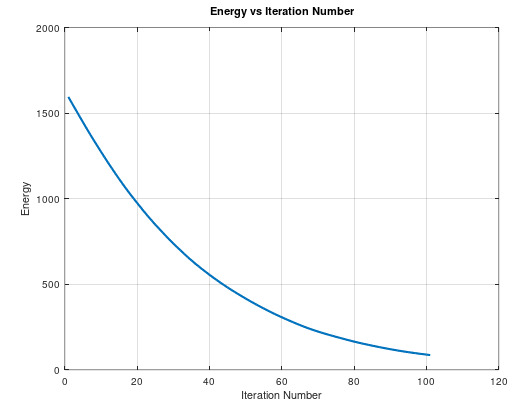}   \\
         \hline
        Fused Sphere (SNR $43.2$ dB)& \includegraphics[scale=0.5]{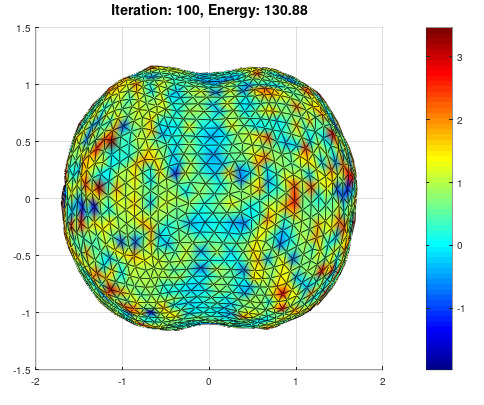} & \includegraphics[scale=0.5]{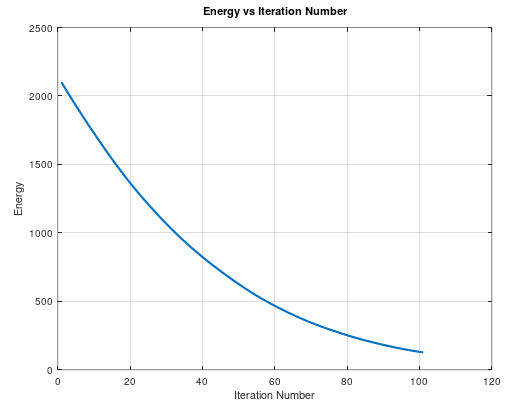}   \\
         \hline
         Cylinder (SNR $42$ dB) & \includegraphics[scale=0.5]{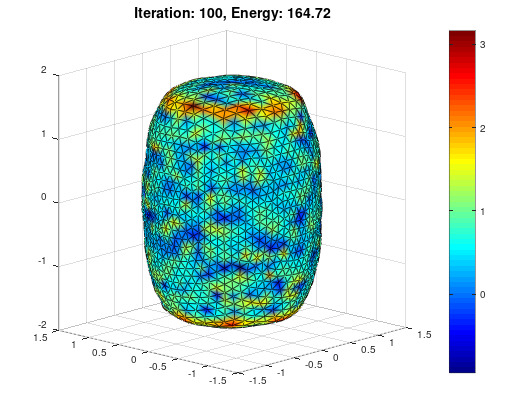} & \includegraphics[scale=0.5]{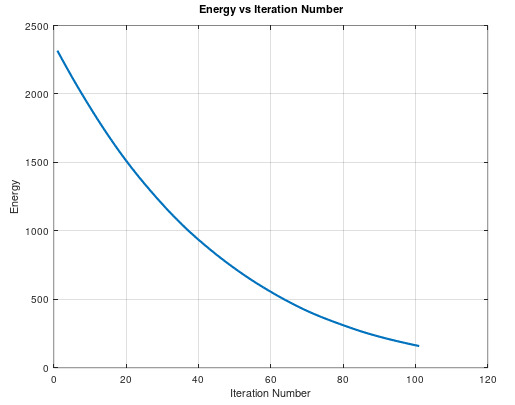}    \\
         \hline
          \end{tabular}}
\caption{Reconstruction of  triangulated surface of phantoms from level sets using shape gradient with tangential term from uniform noisy data.}
    \label{tab:Cylindernoise}
\end{table}

\begin{table}[H]
    \centering
     \resizebox{\linewidth}{!}{%
\begin{tabular}{|c|c|c|c|}
  \hline
  \multirow{2}{*}{Phantoms} & \multicolumn{3}{|c|}{Marching Cube}  \\ \cline{2-4}
                     & Without Noise &  Gaussian Noise & Uniform Noise  \\ \hline
  Unit Sphere  &  \includegraphics[scale=0.25]{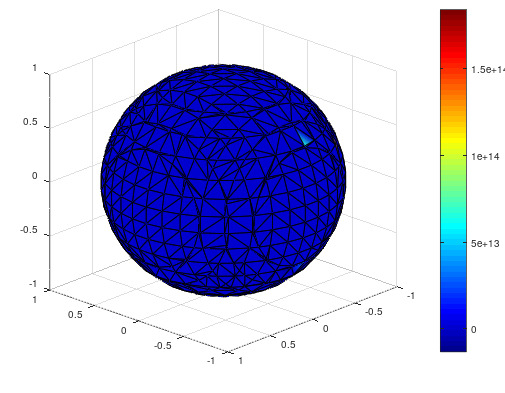} & \includegraphics[scale=0.3]{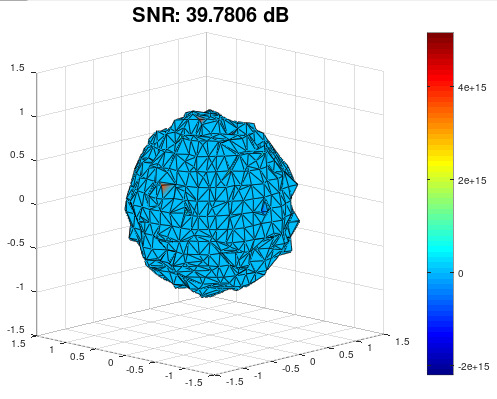}  &  \includegraphics[scale=0.3]{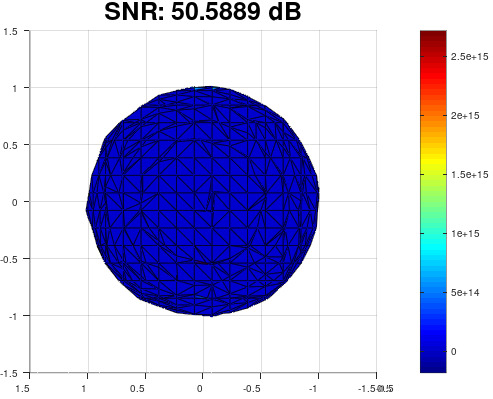} \\ \hline
   Ellipsoid  & \includegraphics[scale=0.25]{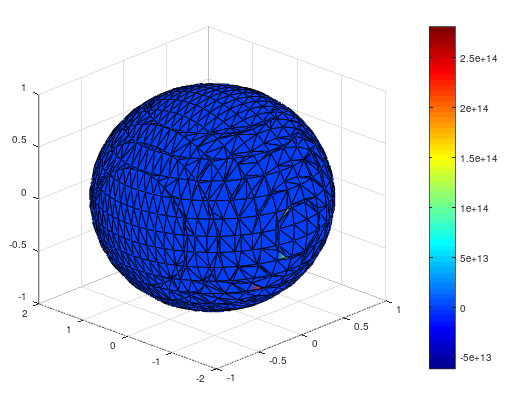} & \includegraphics[scale=0.3]{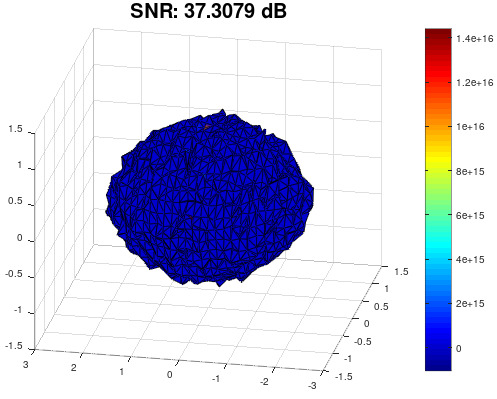} &  \includegraphics[scale=0.3]{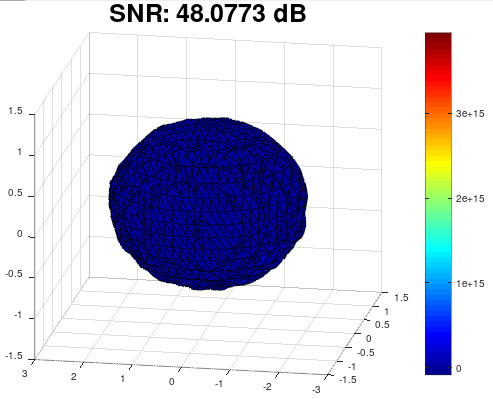} \\ \hline
  Fused Sphere  & \includegraphics[scale=0.25]{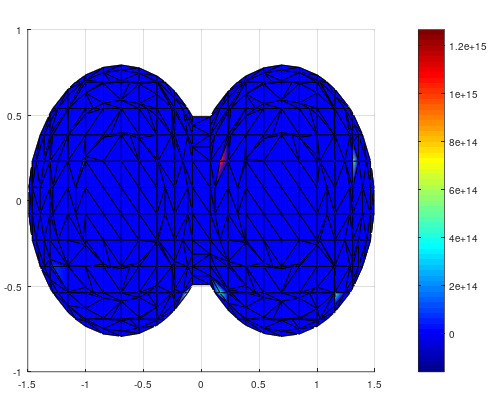} & \includegraphics[scale=0.3]{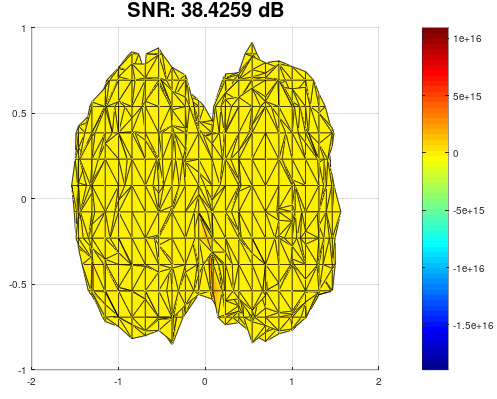}&  \includegraphics[scale=0.3]{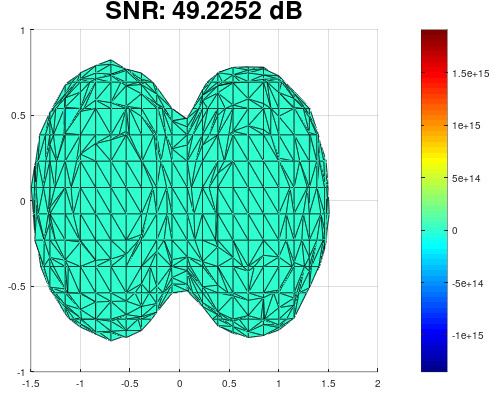}  \\ \hline
  Cylinder  & \includegraphics[scale=0.25]{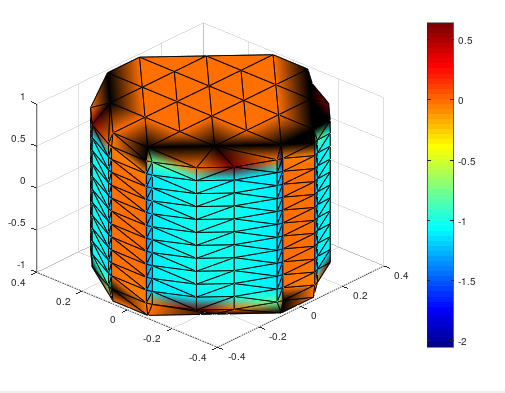} & \includegraphics[scale=0.3]{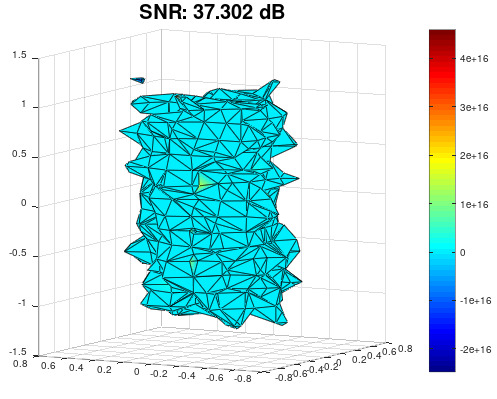} &   \includegraphics[scale=0.25]{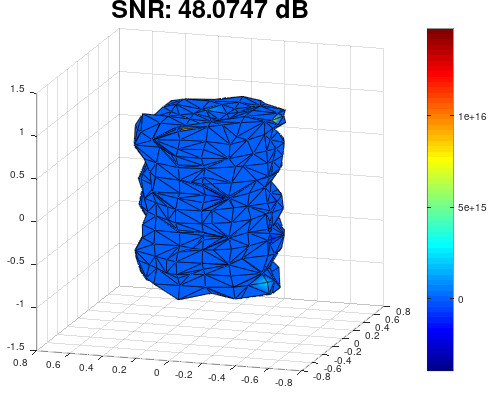} \\ \hline
\end{tabular}}
\caption{Reconstruction of phantoms using the Marching Cube Method. Third and fourth column represent phantoms with the Gaussian and uniform noise data respectively.}
    \label{tab:marching_cube}
\end{table}

\begin{table}
    \centering
     \resizebox{\linewidth}{!}{%
\begin{tabular}{|c|c|c|c|}
  \hline
  \multirow{2}{*}{Phantoms} & \multicolumn{3}{|c|}{DistMesh}  \\ \cline{2-4}
                     & Without Noise &  After First Few Iterations & Further Later  \\ \hline
  Unit Sphere  &  \includegraphics[scale=0.25]{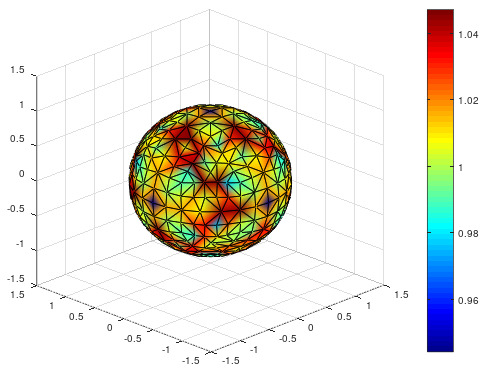} & \includegraphics[scale=0.3]{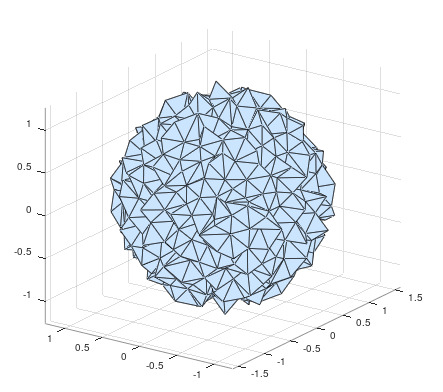}  &  \includegraphics[scale=0.3]{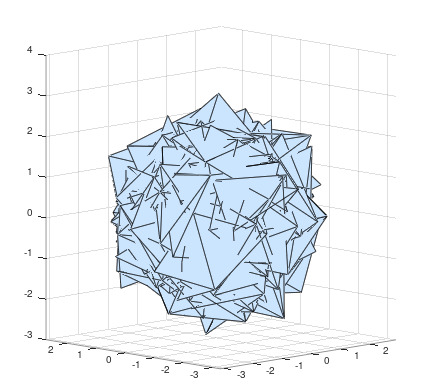} \\ \hline
   Ellipsoid  & \includegraphics[scale=0.25]{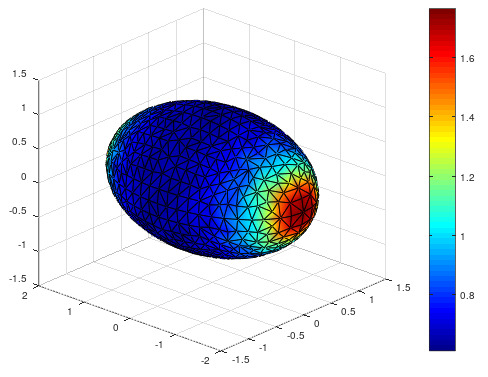} & \includegraphics[scale=0.3]{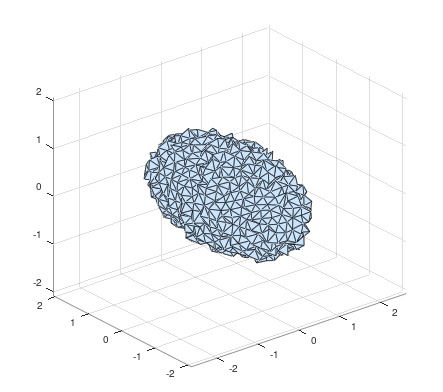} &  \includegraphics[scale=0.3]{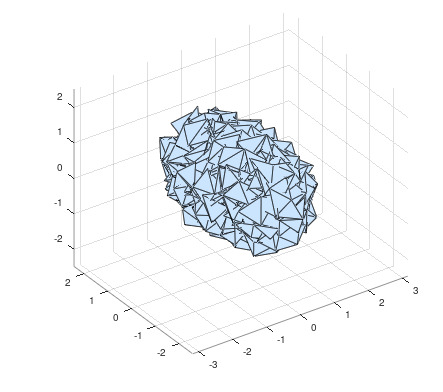} \\ \hline
  Fused Sphere  & \includegraphics[scale=0.25]{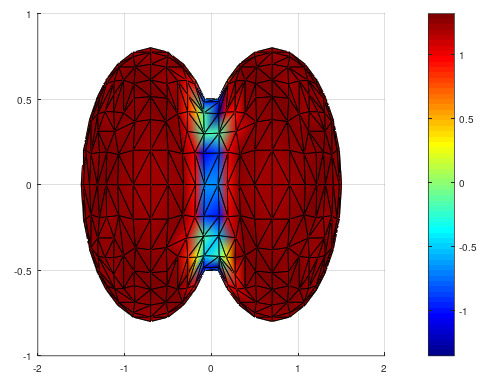} & \includegraphics[scale=0.3]{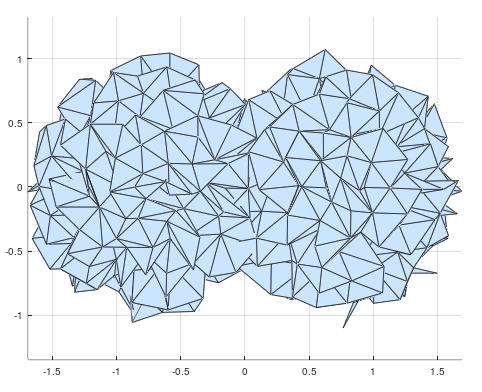}&  \includegraphics[scale=0.3]{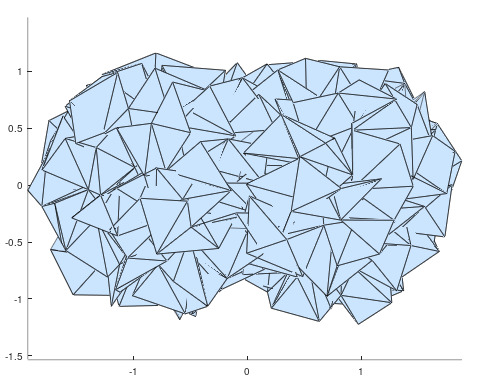}  \\ \hline
  Cylinder  & \includegraphics[scale=0.25]{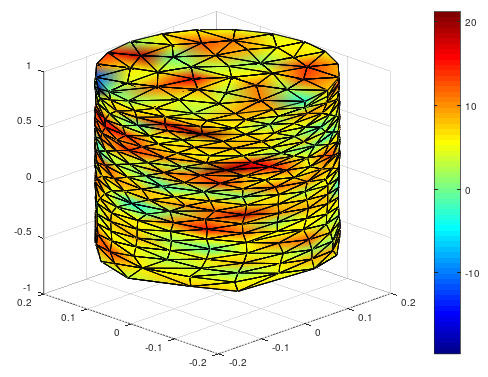} & \includegraphics[scale=0.3]{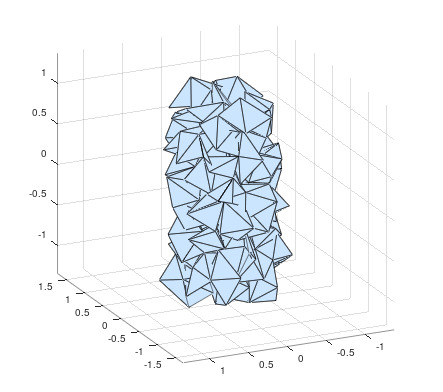} &   \includegraphics[scale=0.25]{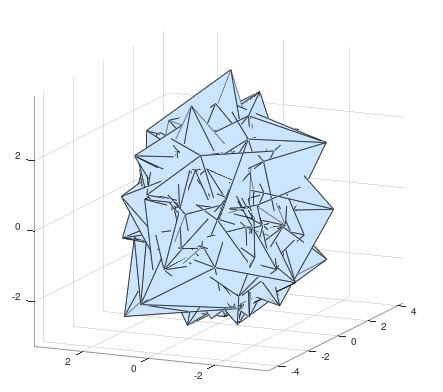} \\ \hline
\end{tabular}}
\caption{Reconstruction of phantoms using the DistMesh Method. The second column represent without noise, third and fourth column represent phantoms with noise respectively.}
    \label{tab:Distmesh}
\end{table}

The study successfully demonstrated that our mathematical model and algorithm effectively reconstruct 2D simple manifolds from level set data using shape gradient. Tests on shapes like the Sphere, Ellipsoid, Fused Sphere, and Cylinder showed consistent strong performance. Adding the tangential term notably improved surface smoothness, as shown by comparisons of trials with and without it. The model also proved robust when tested with noise, maintaining accuracy and effectiveness. Compared to the Marching Cubes Method (See Table \ref{tab:marching_cube}), our approach was not only more accurate in reconstructing 2D phantoms but also better at handling noisy data. Moreover, the energy graph for each trial smoothly converged to zero. These results highlight the model’s robustness and versatility, indicating its potential for precise manifold reconstruction in noisy conditions

\section{Analysis of the computational outcomes}
\textbf{Sphere}s are characterized by their constant curvatures property at each point on the surface, whether it be Mean curvature or Gaussian curvature. The initial surface converges seamlessly to the unit sphere in both cases where $\beta$ is zero or nonzero. In the latter case, the variation in curvature is smaller than the former due to tangential smoothness involvement.\\
 	Furthermore, additive Gaussian noise is added to the level set of the unit sphere to result in an SNR of $44.5$ dB .The trials showed a smooth and close to spherical, but with some regions highlighted by larger positive Gaussian curvatures, possibly due to rare but high-strength noise spikes of Gaussian origin. There is further room for the energy to decay if the evolution continues. 
	Both MCM, and DistMesh were significantly suffered by Gaussian noise. MCM resulted in a very non-smooth surface under uniform noise, whereas DistMesh performed numerically unstable in this case. In the case of uniform noise our method outperformed the other two methods, which can be seen from the curvature profiles.\\
	
	\textbf{Ellipsoid}s are known for their varying positive curvature within a bounded interval. This simple geometric shape is a useful phantom for its gentle variability in curvature; it aids in testing the basic functionality of a reconstruction method.
The trials were done on an ellipsoid with a major axis of 2 units and two minor axes of $1$ unit each.  This should result in Gaussian curvatures $(G)$ to lie in between $\frac{1}{4}$  and $4$ , according to the formula $G(x,y) = \frac{1}{a^2b^2c^2} \left( \frac{1}{(x^2 + y^2/16 + z^2)^2} \right)$. In the absence of noise the reconstruction was good, with  particularly uniform mesh being observed when $\beta$ was on, possibly due to minimized variation between surface elements. But the curvatures were observed to be bounded. However, when Gaussian noise was introduced with SNR $42$ dB, an approximate ellipsoid with many hotspots of high curvature was observed.\\
	The MCM struggled  to give a continuous surface, resulting in excessively high curvatures when no  noise was present. In the noisy case, it failed to approximate the shape. On the other hand, DistMesh did wonderfully in the noiseless case but produced an unstable sequences of meshes when noise seeped in; that too uniform noise.\\

We \textbf{Fused two Spheres} close enough to form a trough at the joint by using intersection of level sets. This phantom expected to has constant curvatures everywhere, but is non-differentiable on the intersection crease, causing ill-defined curvatures on a particular circle. In our numerical experiment the tangential smoothness was switched off (i.e. $\beta=0$) it converged, but with a non-smooth joint of relatively high negative or positive curvatures. With $\beta$ activation on, the trough and its neighbors evolved to smoother mesh as expected, causing smaller negative gradients around the joint. MCM could reconstruct the surface but bad smoothness resulted in extremely high curvatures. DistMesh also could construct well but the resolution was coarse at the joint in the non-noise cases. It is observed that the edge length function of DistMesh was too restrictive to tune and achieve finer resolutions of the mesh. Under both the noises it didn't budge well.\\

Another simpler geometric phantom \textbf{Cylinder}, has a particular feature of zero Gaussian curvature everywhere except at the edges of the disks, in fact non-differentiable. The level set for a finite-cylinder was constructed by combining the level sets of two planes and an infinite cylinder.\\
 Very much the similar evolutionary and numerical phenomenon as that of the fused-spheres case, was observed near cylindrical edges. Inclusion of Gaussian noise badly unstabilized both the MCM and DistMesh surfaces, in contrast to the approximately smoother shape of the manifold generated by the proposed method.
\section{Conclusion}
A shape dependent variational approach to reconstruct a simple 2D manifold in the form of triangulated mesh using level set discrete data is presented. The numerical experiments reveal that, in each iteration, the tangential gradient term plays an important role in the smooth deformation of the surface, resulting in diffeomorphic evolution, while the first terms ensured closeness of the surface to zero-level manifold. The implications of this approach may extend to accurate representation of complex surfaces, such as the cortex of the human brain using MRI data, with potential applications in medical image processing.\\
\section*{Future Work}
The model will be strengthened to help in evolving the surface to capture the highly concave or narrow-pass regions with high negative curvature. This may involve re-engineering the prior level set data just near the regions of interest. Further, to tailor the model for non-simple shapes and explore practically feasible surface initializers is a good immediate challenge to take up this work further, in this direction.  

\section*{Conflict of interest}
The authors declare that they have no known conflicts of interest that could have influenced the research presented in this manuscript. No financial, personal, or professional affiliations have biased the study's design, data collection, analysis, or interpretation. Any potential competing interests have been disclosed transparently in accordance with the journal’s policies.

\section*{Acknowledgment}
The authors thank SRM University, Andhra Pradesh, for providing the fellowship and lab facilities that supported this research. Financial assistance to attend various workshops and acquire journal resources has been instrumental in continuing this study.

\bibliographystyle{apalike}
\bibliography{reference}

\end{document}